\documentclass[twocolumn,english,aps,prb,superscriptaddress]{revtex4-1}
\usepackage[latin9]{inputenc}
\usepackage{amsmath}
\usepackage{amssymb}
\usepackage{graphicx}
\usepackage{esint}
\usepackage{array}
\usepackage{color}
\usepackage[usenames,dvipsnames]{xcolor}

\makeatletter
\@ifundefined{textcolor}{}
{%
 \definecolor{BLACK}{gray}{0}
 \definecolor{WHITE}{gray}{1}
 \definecolor{RED}{rgb}{1,0,0}
 \definecolor{GREEN}{rgb}{0,1,0}
 \definecolor{BLUE}{rgb}{0,0,1}
 \definecolor{CYAN}{cmyk}{1,0,0,0}
 \definecolor{MAGENTA}{cmyk}{0,1,0,0}
 \definecolor{YELLOW}{cmyk}{0,0,1,0}
}

\newcommand{\rs}{\rm\scriptscriptstyle}

\makeatother

\usepackage{babel}
\begin{document}

\title{Bosonic topological insulator intermediate state in the superconductor-insulator transition}

\author{M.\,C.\,Diamantini}

\affiliation{NiPS Laboratory, INFN and Dipartimento di Fisica e Geologia, University of Perugia, via A. Pascoli, I-06100 Perugia, Italy}

\author{A.\,Yu.\,Mironov}

\affiliation{A.\,V.\,Rzhanov Institute of Semiconductor Physics SB RAS, 13 Lavrentjev Avenue, Novosibirsk, 630090 Russia}
\affiliation{Novosibirsk State University, Pirogova str. 2, Novosibirsk 630090, Russia}

\author{S.\,V.\,Postolova}

\affiliation{A.\,V.\,Rzhanov Institute of Semiconductor Physics SB RAS, 13 Lavrentjev Avenue, Novosibirsk, 630090 Russia}
\affiliation{Institute for Physics of Microstructures RAS, GSP-105, Nizhny Novgorod 603950, Russia}

\author{X.\,Liu}
\affiliation{Department of Physics, Harvard University, Cambridge, Massachusetts 02138, USA}

\author{Z.\,Hao}
\affiliation{Department of Physics, Harvard University, Cambridge, Massachusetts 02138, USA}

\author{D.\,M.\,Silevitch}

\affiliation{Division of Physics, Mathematics, and Astronomy, California Institute of Technology, Pasadena, CA 91125, USA}

\author{Ya.\,Kopelevich}

\affiliation{Universidade Estadual de Campinas-UNICAMP, Instituto de F��sica ``Gleb Wataghin"/DFA Rua Sergio Buarque de Holanda, 777, Brasil}

\author{P.\,Kim}
\affiliation{Department of Physics, Harvard University, Cambridge, Massachusetts 02138, USA}

\author{C.\,A.\,Trugenberger}

\affiliation{SwissScientific Technologies SA, rue du Rhone 59, CH-1204 Geneva, Switzerland}

\author{V.\,M.\,Vinokur}

\affiliation{Materials Science Division, Argonne National Laboratory, 9700 S.
Cass. Ave, Lemont, IL 60437, USA.}
\affiliation{Consortium for Advanced Science and Engineering (CASE)
University of Chicago
5801 S Ellis Ave
Chicago, IL 60637}

\begin{abstract}
A low-temperature intervening metallic regime arising in the two-dimensional superconductor-insulator transition challenges our understanding of electronic fluids.  Here we develop a gauge theory revealing that this emergent anomalous metal is a bosonic topological insulator where bulk transport is suppressed by mutual statistics interactions between out-of-condensate Cooper pairs and vortices and the longitudinal conductivity is mediated by 
symmetry-protected gapless edge modes.
We explore the magnetic-field-driven superconductor-insulator transition in a niobium titanium nitride device and find marked signatures of a bosonic topological insulator behavior of the intervening regime with the saturating resistance. The observed superconductor-anomalous metal and insulator-anomalous metal dual phase transitions exhibit quantum Berezinskii-Kosterlitz-Thouless criticality in accord with the gauge theory.
\end{abstract}
\maketitle


An anomalous metallic regime intervening between the superconductor and insulator has been reported in a wide variety of two-dimensional electronic systems experiencing the superconductor-to-insulator transition (SIT)\,\cite{haviland,jaeger,paalanen,Zant1992,Mason1999,Markovic1999,Kapitulnik2008,Bollinger2011,Eley2012,Allain2012,Han2014,Couedo2016,Kapitulnik2017,Park2017,marcus}, and is often referred to as ``Bose metal"\,\cite{Das1999}. Despite decades of dedicated studies\,\cite{goldman,Kap2019}, its nature
remains unclear and poses a challenge to our understanding of electron fluids. The very existence of a 2D metal is at odds with the 2D orthodoxy, as conventional theories expect a
direct quantum SIT with no intermediate metallic phase.

\begin{figure}[t!]
	\includegraphics[width=9cm]{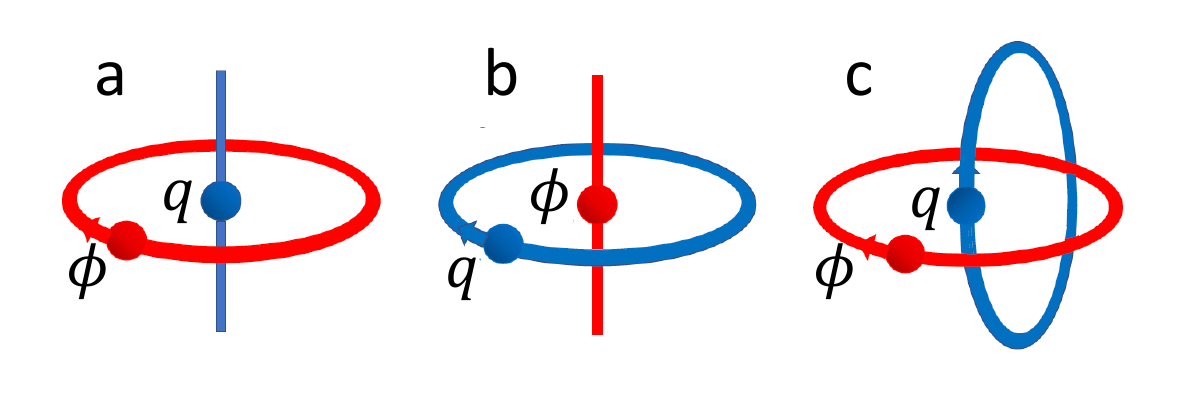}
	\caption{\textbf{Mutual statistics interactions.}
		\textbf{a:} A charge (blue ball) evolves along the Euclidean time axis and is encircled by a vortex (red). The two trajectories are topologically linked since one cannot decouple the charge trajectory from the circle without breaking it. This linking encodes the Aharonov-Casher effect. \textbf{b:} A dual situation in which a vortex (red ball) evolves along the Euclidean time direction and is encircled by a charge. This linking represent the Aharonov-Bohm effect. \textbf{c} The linked charge-anticharge and vortex-antivortex fluctuations representing coupled Aharonov-Bohm-Casher interactions which are encoded in the Chern-Simons term in the action.}
	\label{Fig1}
\end{figure}

Yet a gauge theory of Josephson junction arrays (JJA) at $T$$=$$0$\,\cite{dst} predicted a metallic phase intervening between the superconductor and superinsulator. Below, building on our field theory of the SIT\,\cite{dtv}, we extend the approach of\,\cite{dst} to finite temperatures and construct a gauge theory of the Bose metal. We unravel the fundamental role of infrared-dominant Aharonov-Bohm-Casher (ABC) mutual statistics interactions determining the SIT phase structure. We report transport measurements on niobium titanium nitride (NbTiN) and van der Waals heterostructures of twisted double bilayer graphene (TDBG) films offering strong support to the proposed picture. We show that  strong quantum fluctuations prevent Bose condensation of both vortices and Cooper pairs (CP). The mutual statistics interactions, see Fig.\,\ref{Fig1}, between them induce a gap in their  fluctuation spectrum, quantified by the Chern-Simons mass, $m_{\rs CS}$\,\cite{jackiw},
preventing bulk transport. The longitudinal conductance is mediated by symmetry-protected U(1)$\rtimes\mathbb{Z}_2^{\rm T}$ edge modes, where $\mathbb{Z}_2^{\rm T}$ denotes time-reversal symmetry. Hence the Bose metal realizes the long sought\,\cite{Balents2017}  bosonic topological insulator.


\begin{figure*}[t!]
	\includegraphics[width=17cm]{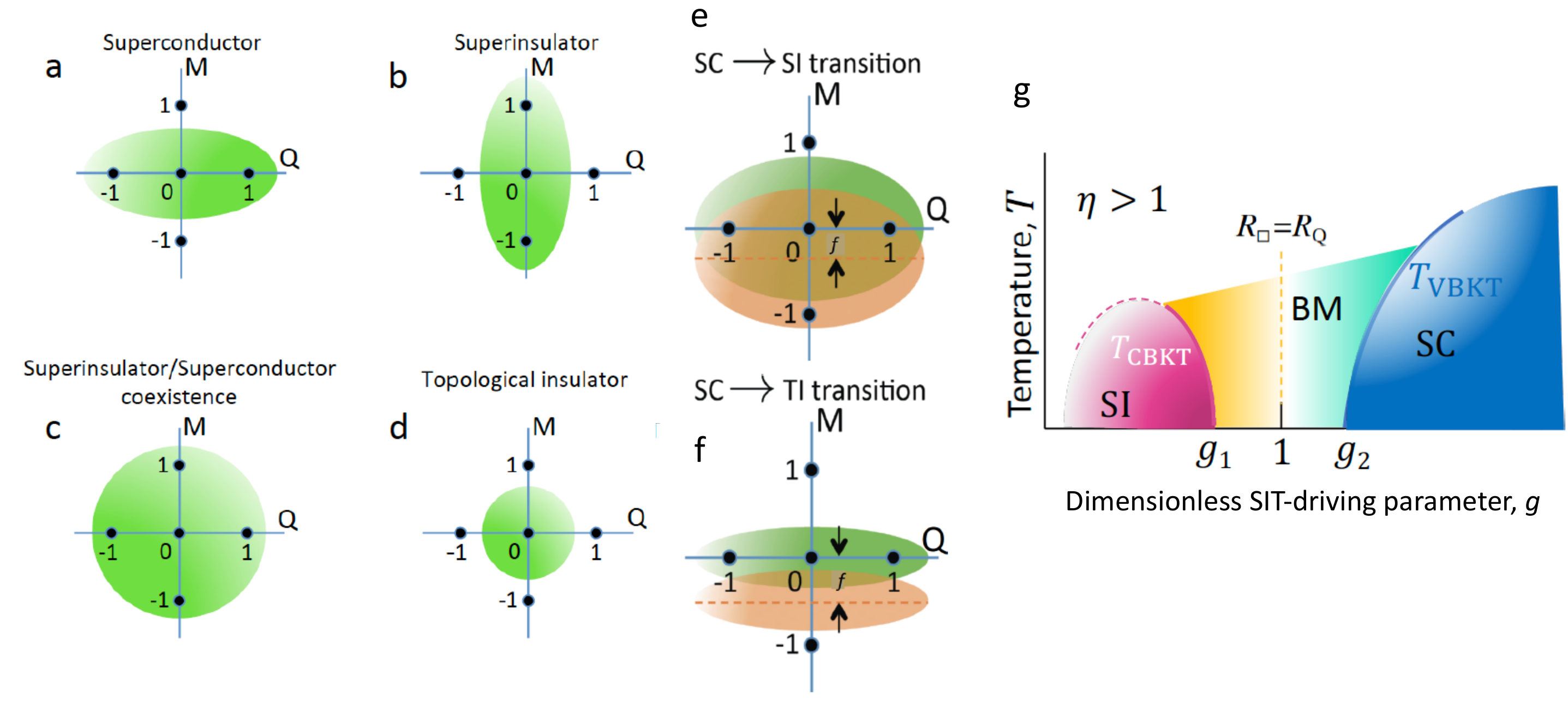}
	\vspace{-0.5cm}
	\caption{\textbf{Phase structure of the SIT.} 
		\textbf{a:} 
		Superconductor: Strings with electric quantum numbers condense. \textbf{b:} Superinsulator: Strings with magnetic quantum numbers condense. \textbf{c:} Coexistence of long electric and magnetic strings
		near the first-order direct transition from a superconductor to a superinsulator. 
		\textbf{d:} Bosonic topological insulator/Bose metal: all strings are suppressed by their high self-energy. \textbf{e:} Phase transitions induced by an external magnetic field: Direct, magnetic-field-induced transition from a superconductor to a superinsulator for $\eta <1$.
		\textbf{f:}\,Magnetic-field-driven transition from a superconductor to a topological insulator for $\eta > 1$.
		\textbf{g:} Schematic phase diagram for strong quantum fluctuations, $\eta>1$ (not to scale), near the SIT. The Bose metal/topological insulator state (BM) is separated from the superconducting (SC) and superinsulating (SI) states by vortex- and charge-BKT transitions respectively (magenta and blue solid lines respectively). In the vicinity of the $T_{\rs VBKT}$ line, $R_{\rs\square}$$<$$R_{\rs Q}$ and the system exhibits metallic behavior which crosses over smoothly into the thermally activated insulating behavior upon decreasing $g$. At the dotted orange line, $R_{\rs\square}$$=$$R_{\rs Q}$ and the resistance keeps increasing, $R_{\rs\square}$$>$$R_{\rs Q}$, towards the $T_{\rs CBKT}$ line. The yellow strip depicts the domain of thermally activated resistance, which crosses over to BKT critical behavior on approach to $T_{\rs CBKT}$. The SIT-driving values $g_1$ and $g_2$ that mark superinsulator-Bose metal and superinsulator-Bose metal quantum transitions satisfy the duality relation $g_2=1/g_1$. The parameter $g$ can be either the dimensionless conductance of the film, or magnetic field, or gate voltage in the gate-driven SIT.
	}
	\label{Fig2}
\end{figure*}


\section*{Phase structure of the SIT}~~
We consider a two-dimensional superconducting film in the vicinity of the SIT at temperatures $T\leqslant T_{\mathrm c0}$, where $T_{\mathrm c0}$ is the mean-field temperature of formation of  Cooper pairs with the infinite lifetime. The film
harbors interacting elementary excitations, Cooper pairs and vortices,  with characteristic energies $e_{\mathrm q}^2=4e^2/\ell$ and $e_{\mathrm v}^2=\Phi_0^2 /\lambda_{\perp}$, where $e$ is the electron charge (we use natural units, $c=1$, $\hbar =1$, restoring physical units when necesssary),  $\Phi_0=\pi /e$ is the flux quantum,
the ultraviolet (UV) cutoff of the theory is defined as  $\ell\simeq{\textrm{min}\{d,\xi\}}$, with $d$ and $\xi$ being  the thickness of the film and the superconducting coherence length, respectively,  and $\lambda_{\perp} = \lambda_{\rs L}^2/d$ is the Pearl length of the film, with $\lambda_{\rs L}$ being the London penetration depth of the bulk material. Mapping to JJA is achieved via the $e_{\mathrm q} \to 4E_{\rs C}$, $e_{\mathrm v} \to 2\pi^2 E_{\rs J}$ replacement, where $E_{\rs C}$ and $E_{\rs J}$ are the charging and Josephson coupling energies of the array, respectively. Since $E_{\rs J}\sim g\Delta$, where $\Delta$ is the gap in a superconducting granule in the JJA and $g$ is the dimensionless tunneling conductance between the adjacent granules, the model implicitly takes into account dissipation effects.

The behavior of an ensemble of interacting vortices and CP near the SIT is governed by the free energy which we derive
from the mixed Chern-Simons action\,\cite{dtv}, %
explicitly accounting for the ABC effects\,\cite{wilczek,jackiw}. To that end we integrate out the gauge fields and
arrive at the free energy of a  system of strings carrying electric and magnetic quantum numbers $Q$ and $M$ and representing the Euclidean trajectories of charges and vortices on a lattice of spacing $\ell$, see Appendix:
${\cal F}=\left({Q^2}/{g}+gM^2-{1}/{\eta} \right) \mu_{\mathrm e} \eta N$,
where the string length $L=N\ell$ and $g$$=$$e_{\mathrm v}/e_{\mathrm q}$$=$$(\pi/e^2) \sqrt{\ell/ \lambda_{\perp}}$ is the dimensionless tuning parameter with $g$$=$$g_{\mathrm c}$=$1$ corresponding to the SIT.
In experiments on films experiencing the disorder-driven SIT, the tuning parameter is the  dimensionless conductance, $g=R_{\rs Q}/R_0$, where $R_{\rs Q}=h/4e^2$ is the quantum resistance for CP, $h$ is the Planck constant, and
$R_0$ is the sheet resistance of the film measured at predefined standard conditions.  The parameter $g$ describes thus the tuning of the SIT not only by regulating disorder, but implicitly also by varying dissipation\,\cite{fazio} (in JJA $g=\sqrt{(\pi^2/2)(E_{\rs J}/E_{\rs C})}$).
The dimensionless parameter $\eta$ describes the strength of quantum fluctuations. Near the SIT, where $e_{\mathrm q}$$\approx$$e_{\mathrm v}$, $\eta$ acquires the form (in physical units)
\begin{equation}
\eta = {1\over \alpha}  {\pi^2 \ell \over \mu_e \lambda_{\perp}} {\ G\left( \pi \ell \over \alpha \lambda_{\perp} \right)} \,,
\label{eta}
\end{equation}
where $\alpha = e^2/(\hbar c) \approx 1/137$ is the fine structure constant, $\mu_e $ is the positional entropy (see SI), and $G$ is the diagonal element of the 3D Green function describing electromagnetic interactions screened by the CS mass, see\,SI.  Identifying the UV cutoff with the superconducting coherence length $\xi$, yields the geometric factor as $d/(\kappa\lambda_{\rs L})$ where $\kappa=\lambda_{\rs L}/\xi$ is the Landau-Ginzburg parameter.

\begin{figure*}[t!]
	\includegraphics[width=21cm]{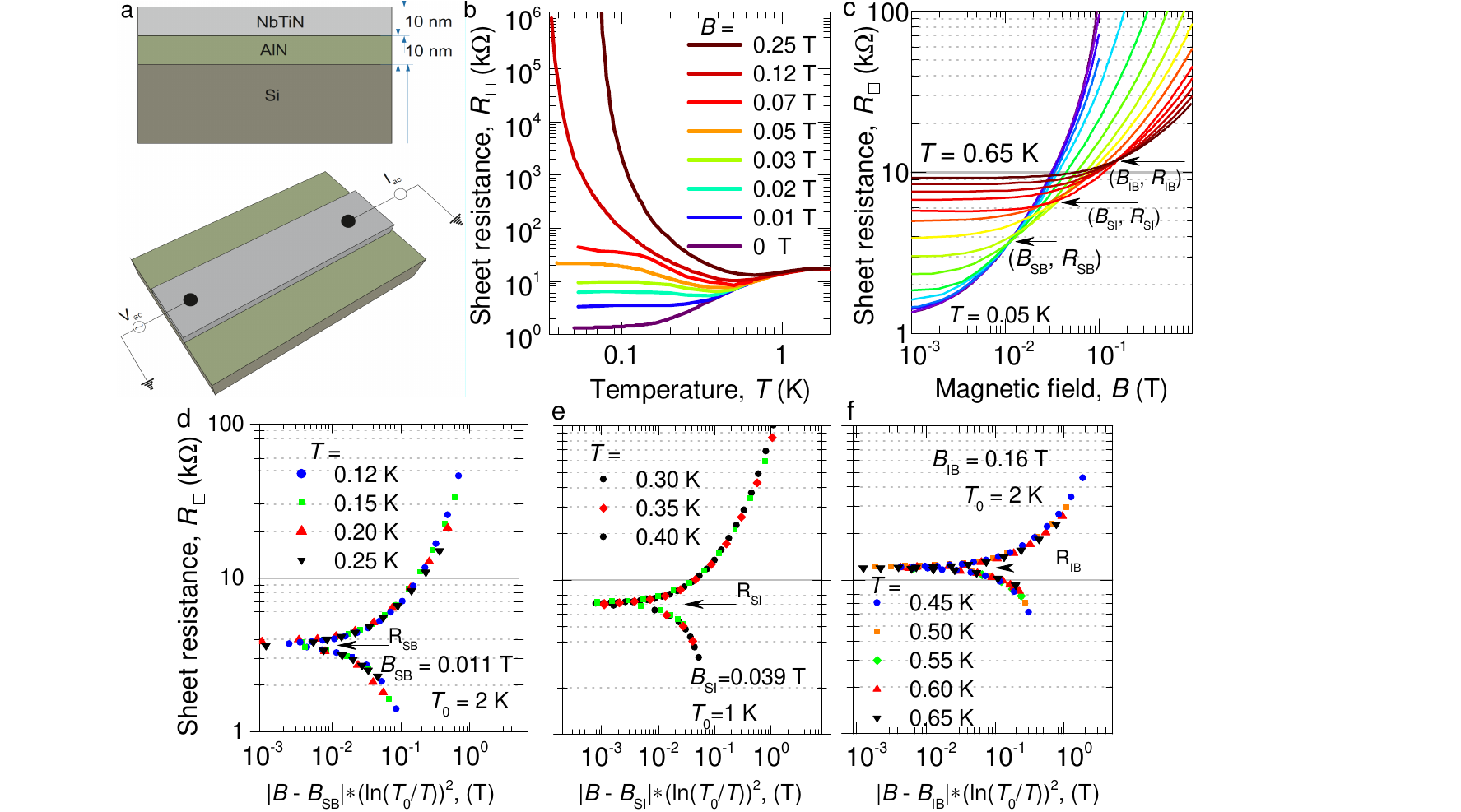}
	\vspace{-0.5cm}
	\caption{\textbf{Temperature and magnetic field dependencies and quantum BKT scaling of $R_{\rs\square}(T,B)$ in the vicinity of the magnetic field-driven SIT.}
		\textbf{a:} The device scheme and a setup for the two-terminal resistance measurements sketch.
		\textbf{b:} Sheet resistance, $R_{\rs \square}$, as function of temperature, $T$, spanning the the anomalous metallic regime in the magnetic field range from 0\,T to
		0.25\,T.
		The residual resistance grows with decreasing $g$ and becomes $R_{\rs\square}=R_{\rs Q}$ at $g\simeq 1$.
		\textbf{c:} Sheet resistance, $R_{\rs\square}$, as function of $B$ for
		different temperatures exhibits three crossing points (marked by arrows) at $B_{\rs SB}=0.011$\,T, $B_{\rs IB}=0.16$\,T, and $B_{\rs SI}=0.039$\,T, corresponding to superconductor-BM, BM-superinsulator, and the SIT transitions, respectively. The crossing points  satisfy the duality relation  $B_{\rs SB}/B_{\rs SI}=B_{\rs SI}/B_{\rs IB}$ with great accuracy.
		\textbf{d:} Quantum BKT scaling near the $B_{\rs SB}$ critical point.
		\textbf{e:} The BKT scaling for the remnant of the SIT.
		\textbf{f:} Quantum BKT scaling near the $B_{\rs IB}$ critical point.
	}
	\label{Fig3}
\end{figure*}

Bose condensation of charges and/or vortices means proliferation of strings of an arbitrary size and occurs if ${\cal F}$ is negative, i.e. if
\begin{equation}
Q^2/g + g M^2<1/\eta\,.
\label{condition}
\end{equation}
The phase emerging at particular values of $g$ and $\eta$ is determined by the geometric condition that the nods on a square lattice of integer electric and magnetic charges, $\{Q,M\}$, fall within the interior of an ellipse with semi-axes  $r_{\rs Q} = (g/\eta)^{1/2}$ and $r_{\rs M}=1/(g\eta)^{1/2}$, see Fig.\,\ref{Fig2}(a--d). The Bose metal emerges if none of the condensates can form i.e. if simultaneously
\begin{equation}
g/\eta>1 \,\,\textrm{and}\,\,g\eta>1\,.
\label{BM}
\end{equation}
This relation resolves the enigma of why some materials exhibit a direct SIT while others go through the intermediate Bose metal phase. The direct SIT at $g=1$ corresponds to $\eta<1$; tuning $g$, one crosses the region $\eta<g<1/\eta$, where both vortex and Cooper pair condensates coexist, i.e. the direct SIT is a first-order quantum transition. The Bose metal phase opens up for $\eta >1$ and is favored by thicker films and for materials with low $\kappa$. Its domain is delimited by the lines $g$$=$$\eta$ and $g$$=$$1/\eta$.  As we show below, these lines represent quantum
Berezinskii-Kosterlitz-Thouless\,\cite{ber,kos} (BKT) transitions, and $g$$=$$1$, $\eta$$=$$1$ is a quantum tri-critical point.

To describe the magnetic-field-driven SIT in systems with  $g\approx 1$ i.e. which are already on the brink of the SIT, we introduce the frustration factor $f=B/B_{\rs \Phi}$, where $B_{\rs \Phi}$ is the magnetic field corresponding to one flux quantum $\Phi_0=\pi /e$ per unit cell. Then the external magnetic field shifts $M\to M+f$ and modifies the condensation conditions, see Fig.\,\ref{Fig2}e,f. Setting $g=1+\epsilon$, with $\epsilon\ll 1$, one finds for a direct SIT at $\eta<1$, $f_{\rm c} = (1/2)(g^2-1)\approx\epsilon$.
At $\eta>1$, but still close to the tri-critical point, one sets $g=\eta+\epsilon$ and obtains for the superconductor-BM transition $f_{c}=\sqrt{\epsilon}/\eta^{3/2}=(g-\eta)^{1/2}/\eta^{3/2}$ see\,SI for details.


\section*{The nature of the intermediate Bose metal}~~
The response of a BM to an applied field is determined by its effective electromagnetic action, obtained by integrating out the effective gauge fields in the general Chern-Simons action (see Appendix), which (without any loss of generality) assumes the simplest form in the relativistic case 
\begin{equation}
S_{\rm eff} \left( A_{\mu} \right) = \frac{g }{2} \left( \frac{\bar{q} e}{2\pi} \right)^2 d \int d^3 x \ A_{\mu} \left( -\delta_{\mu \nu} \nabla^2 + \partial_{\mu} \partial_{\nu} \right) A_{\nu} \,,
\label{bulkaction}
\end{equation}
where $A_{\mu}$ is the external electromagnetic potential, the dimensionless charge unit $\bar{q}=2$ for CP and the renormalized effective charge
$\bar{q} e_{\rm eff} = \bar{q} e \sqrt{g}$. Accordingly, the charge current $j_{\mu}$ is found as  $j_{\rm ind}^{\mu} = (\delta /\delta A_{\mu}) S_{\rm eff} \left( A_{\nu} \right)=
g \left( {\bar{q} e/ 2\pi} \right)^2 d \ \partial_{\nu} F^{\mu \nu} $, with $F_{\mu \nu} = \partial_{\mu} A_{\nu} - \partial_{\nu} A_{\mu}$. One sees that only the derivatives of the external fields, but not their constant parts, induce a current. Therefore, in the bulk, both the longitudinal and the quantum Hall components of the linear conductance vanish at $T=0$.%

However, the Chern-Simons effective action is not invariant under gauge transformations not vanishing on the sample boundaries. To restore the gauge invariance, one has to add edge chiral bosons\,\cite{flore}. This operation is an exact analogue to the description of the edge modes in the quantum Hall effect\,\cite{wen}. The resulting edge action gives rise to the equation of motion, see Appendix:
\begin{equation}
v_{\mathrm b} \rho = \frac{\bar{q} e_{\rm eff}}{2\pi} V\ ,
\label{response}
\end{equation}
where $v_{\mathrm b}$ is the velocity of the edge modes and $V$ is the applied voltage. Using $I=\bar{q}e_{\rm eff}v_{\mathrm b}\rho$, one finds the longitudinal sheet resistance
\begin{equation}
R_{\rs\square}\equiv {V\over I} = {R_{\rs Q}\over g} \,,
\label{R}
\end{equation}
which is in a full concert with the early elegant charge-vortex duality arguments\,\cite{fisher1990,fazio} leading to $R_{\rs{\square}}=R_{\rs Q}$\,\cite{fisher1990,fisher1992} at the SIT. In 
experiments, $R_{\rs\square}$ may well deviate from $R_{\rs Q}$ upon departure from the SIT, but the standard SIT scaling analysis
yields the convergence of $R_{\square}(T\to 0)$ to  $R_{\rs Q}$\,\cite{paalanen,jaeger,marcus}  upon approach to the presumed quantum critical point.
At zero temperature and $\eta$$>$$1$, the Bose metal forms for  between the points $g=1/\eta$ and $g=\eta$. Hence its sheet resistance is lower than the quantum resistance, $R_{\rs\square} < R_{\rs Q}$, on the superconducting side, $g > 1$ and larger than the quantum resistance, $R_{\rs\square} > R_{\rs Q}$, on the insulating side, $g< 1$, with equality achieved at $g=1$. Duality is realized in the form $R_{\rs Q}/R_{\rs\square}$$ \leftrightarrow$$R_{\rs\square}/R_{\rs Q}$ when $g\leftrightarrow 1/g$, generalizing the universality arguments of\,\cite{fisher1990, fisher1992, fazio} onto the Bose metal.

Bulk transport suppressed by the topological gap and ballistic symmetry-protected edge modes are the hallmark of topological insulators. In our case, while the flux quantum is  $\pi/e$, the charge is carried by bosonic excitations of charge $2e$. The BM is thus an integer bosonic topological insulator with edge modes protected by the U(1)$\rtimes\mathbb{Z}_2^{\rm T}$ symmetry. This is one of the generic integer topological phases recently classified in\,\cite{vis,Senthil2013}.
The quantum fluctuations parameter $\eta$, the suppression of bulk conductances by the topological CS mass $m_{\rm CS}$, and equations (\ref{response},\ref{R}) for the longitudinal resistivity mediated by the U(1)$\rtimes\mathbb{Z}_2^{\rm T}$-symmetry protected gapless edge states are the central results of our work.

Shown in Fig.\,\ref{Fig2}g is a general phase diagram comprising the BM near the SIT, with the tuning parameter $g$ denoting either the conductance, or magnetic field or gate voltage in a proximity array\,\cite{marcus}. The BM forms at $\eta >1$ and occupies the area between the charge- and vortex-BKT transition lines. Quantum transitions between the superinsulator and the BM and between the superconductor and the BM occur at $g=g_1$, and $g=g_2$, respectively. Duality requires $g_2=1/g_1$. By self-duality, the SIT can still be identified as the line $g=1$ on which $R_{\rs\square}(T) =R_Q$ (the dashed line in Fig\,\ref{Fig1}g) although there is no phase transition anymore at this point.


\begin{figure*}[t!]
	\includegraphics[width=19cm]{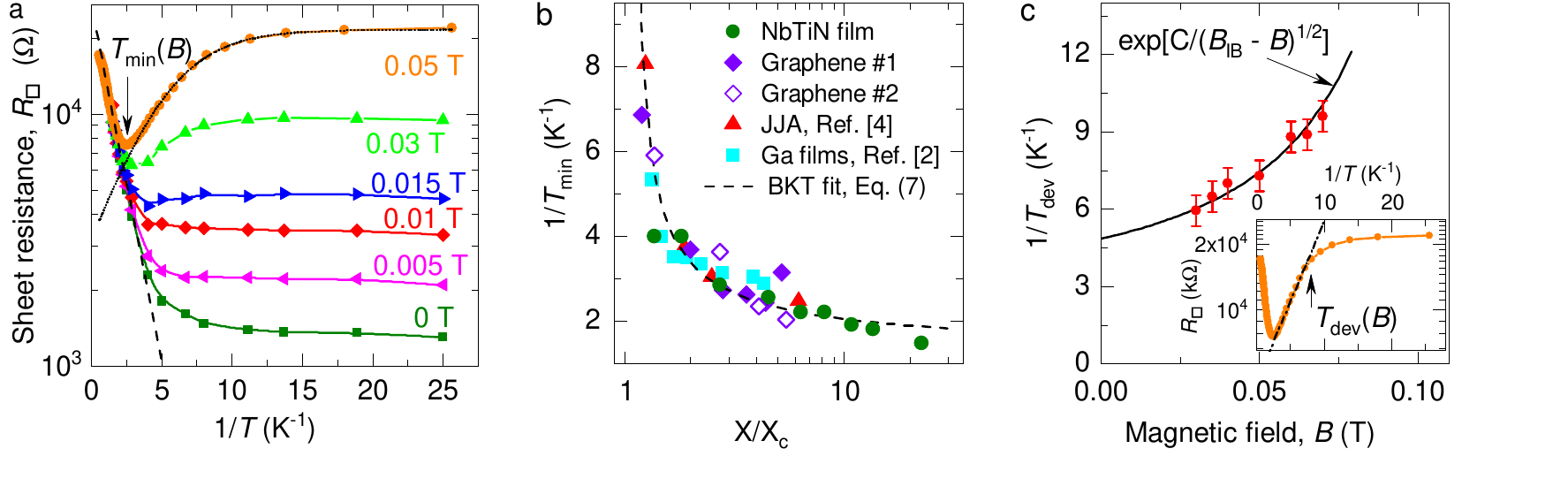}
	\vspace{-0.9cm}
	\caption{\textbf{Sheet resistance vs $1/T$ plots and the BKT scaling for correlation lengths.}
		\textbf{a:} Representative $\log R_{\rs \square}$ vs. $1/T$ plots. The dashed black line depicts $R\propto\exp(-T^*/T)$ dependence. The dotted line describes $R_{\rs\square}(1/T)$ behavior for two parallel resistors comprising $R_{\mathrm b}\propto\exp(T^{**}/T)$ (bulk modes) insulating dependence and constant (edge ballistic modes) behavior, see text.
		\textbf{b:} The BKT scaling of $1/T_{\mathrm{min}}\sim\xi_{\mathrm{corr}}\sim\exp[\mathrm{const}/(X/X_{\mathrm c}-1)^{1/2}]$ of the BKT correlation length near the superconductor-BM quantum BKT transition in different materials driven by either magnetic field, disorder, or gate voltage, demonstrating the universality of the BKT superconductor-BM transition. The circles stand for $T_{\mathrm{min}}(B)$ in NbTiN films, $X/X_{\mathrm c}=B/B_{\rs SB}$; the diamonds mark data for Van der Waals heterostructures of twisted double bilayer graphene (TDBG) 
		where the SIT is tuned by the in-plane magnetic field, $X/X_{\mathrm c}=B_{\rs{||}}/B_c$, $B_c =2.5$\,T is an adjustment parameter and $T_{\mathrm{min}}$ is divided by 35; the triangles mark data for the SIT driven by the frustration factor $f=\Phi/\Phi_0$, with $\Phi$ being the magnetic flux per plaquette in JJA, $f_{\mathrm c}=0.1$ \,\cite{Zant1992}, and squares mark the film thickness driven SIT in granular films\,\cite{jaeger}, for which $X/X_{\mathrm c}=R_{\rs N}/R_{\rs Nc}$, $R_{\rs Nc}=32$\,k$\Omega$, $T_{\mathrm{min}}$ is divided by 16.
		\textbf{c:} The BKT scaling of the correlation length $\sim 1/T_{\mathrm{dev}}$ near the insulator-Bose metal transition, detected by deviation of the resistance from the exponential insulating $\exp(T^{**}/T)$ behavior (the dashed line in the inset).
	}
	\label{Fig4}
\end{figure*}

\section*{Stability of the bosonic topological insulator}~~
To unravel the mechanism preventing the condensation of CP and ensuring stability of the Bose metal, we recall that the particle number operator $N_q$ and the U(1) phase $\varphi$, form a pair of canonically conjugate variables, since $N_{\mathrm q}$  is the generator of global U(1) charge transformations.  Only two symmetry realizations are allowed in infinite systems. Either (i) $N_{\mathrm q}$ is fixed, $\Delta N_{\mathrm q}=\sqrt{\langle \left( N_{\mathrm q} -\bar N_{\mathrm q} \right)^2 \rangle} =0$,
and $\varphi $ is undefined, $\Delta\varphi  = \infty$ so that the U(1) symmetry is linearly realized, or (ii) $N_{\mathrm q}$ does not annihilate the vacuum, $\Delta N_{\mathrm q}$$=$$\infty$, $\varphi $ is fixed, $\Delta \varphi$$=$$0$ and the global U(1) symmetry is spontaneously broken. These two possibilities define the zero-temperature superinsulators and superconductors\,\cite{vinokur2008superinsulator, vinokurAnnals}. However, because of the topological Chern-Simons interactions, the charge and vortex numbers do not obey anymore the Gauss-law constraints generating the two U(1) symmetries, and the third possibility, $\Delta N_{\mathrm q}\neq 0$, $\Delta N_{\mathrm v}\neq 0$, arises.
We find that in the Bose metal, the equal-time quantum correlation functions in the ground state are given by
\begin{eqnarray}
\langle j^0({\bf x}) j^0({\bf y}) \rangle \propto {\rm exp} \left(- \frac{|{\bf x}-{\bf y}|}{\xi_{\mathrm{corr}}(g/g_{1})} \right)\,, \nonumber \\
\langle \phi^0({\bf x}) \phi^0({\bf y}) \rangle \propto {\rm exp} \left(- \frac{|{\bf x}-{\bf y}|}{\xi_{\mathrm{corr}}(g_{2}/g)} \right) \ ,
\label{corrbkt}
\end{eqnarray}
where $j^0$ and $\phi^0$ are the charge and vortex densities, respectively and
$\xi_{\mathrm{corr}} (x) \propto {\rm exp} ( {\rm const}/ \sqrt{|x-1|}) $\ ,
is the BKT\,\cite{ber, kos} correlation length with $x$ set by the quantum coupling constant $g$ and its critical point $g_c$.
Accordingly, the two dual transitions, superconductor-BM and superinsulator-BM are quantum BKT transitions. Another far-reaching implication of Eq.\,(\ref{corrbkt}) is that charges and vortices form an intertwined liquid comprising fluctuating macroscopic islands with typical dimensions $\xi(g/g_1)$ and $\xi(g_2/g)$, respectively. The emergent texture is referred to as the self-induced electronic granularity\,\cite{vinokurAnnals}. The associated characteristic frequency of the BM quantum fluctuations is $\omega = v/\xi_{\rm corr}$. While the exact expression for the correlation length in the ground state is not yet available, it must lie in the interval $\xi < \xi_{\rm corr} < \hbar/(v m_{\rs CS}) $, where the upper bound is the length scale associated with the CS gap to the first excited state. We obtain thus the lower bound $\omega > m_{\rs CS} v^2/\hbar$. Using $m_{\rs CS} = \hbar e_{\mathrm q} e_{\mathrm v} /\pi v$ and $e_{\mathrm v} $$\approx$$e_{\mathrm q}$ at the center of the BM phase, we find $\omega$$ >$$\alpha v/d$. This is the typical frequency associated with the electrostatic energy $\hbar \alpha v/d$ of a Cooper pair. For the NbTiN parameters\,\cite{Mironov2018} $d= 10$\,nm and $v/c = 1/\sqrt{\varepsilon} = 1/\sqrt{800}$, we find $\omega > 7$\,THz.

\vspace{-0.36cm}
\section*{Experiment}~~
Transport measurements are taken on NbTiN 10\,nm thick films prepared by the atomic layer deposition (ALD) technique based on sequential
surface reaction, step-by-step film growth. The films were lithographically patterned into bars, see Fig.\,\ref{Fig3}a,
and resistivity measurements were performed at sub-Kelvin
temperatures in helium dilution refrigerators (see the details of the sample preparation, geometry, measurement technique and characterization
in\,\cite{Mironov2018}). All the resistance measurements were carried out in the linear regime, with the proper electric line filtering. Since the expected frequency of typical quantum fluctuations responsible for BM behaviour exceeds by several orders of magnitude the system of filtration, this is effective for noise elimination but does not affect the relevant physics at higher frequencies. Shown in Fig.\,\ref{Fig3}a is the sketch of the two-terminal setup. Figure\,\ref{Fig3}b presents the log-log plot of the low-temperature part of $R_{\rs\square}(T) $ across the magnetic field-driven SIT. At magnetic fields $B\lesssim 0.04$\,T, $R_{\rs \square}$ saturates at lowest measured temperatures to the magnetic field-dependent value spanning about an order of magnitude in sheet resistance, from  $R_{\rs \square}\simeq 1$\,k$\Omega$ to $R_{\rs \square}\simeq 20$\,k$\Omega$, suggesting metallic behaviour across this range. At fields above $B\simeq 0.011$\,T, the $R_{\rs \square}(T)$ dependence develops a minimum. Above $B\gtrsim 0.04$\,T, the curve shows a trend to an insulating upturn, and, as soon as $B$ exceeds 0.16\,T, $R_{\rs \square}(T)$ exhibits pronounced insulating behaviour. 
Plotting $R_{\rs \square}(B)$ isotherms, see panel\,Fig.\,\ref{Fig3}c, exposes three sequential crossing points $B_{\rs {SB} }=0.011\pm 0.001$\,T,  $B_{\rs SI}=0.039\pm 0.001$\,T, and $B_{\rs IB}=0.16\pm 0.01$\,T.  The corresponing resistances are $R_{\rs SB}=3.63\pm 0.01$\,k$\Omega$, $R_{\rs SI}=6.57\pm 0.01$\,k$\Omega$, and $R_{\rs SI}=11.9\pm 0.1$\,k$\Omega$.

Temperature dependencies of the resistance $R(T)$ of TDBG measured at optimal doping in the presence of the magnetic field, $B_{\rs{||}}$, parallel to the film indicates the field-induced SIT, see\,\cite{Kim2019}, where also the details of fabrication of TDBG devices and measurements protocol can be found. In the intermediate $B_{\rs{||}}$ region $R(T)$ develops a minimum the position of which depends on $B_{\rs{||}}$. As we discuss below, these minima may indicate the possibility of the formation of the bosonic topological insulator state.
\vspace{-0.3cm}
\section*{Discussion and conclusion}~~
We identify the crossing points $B_{\rs SB}$ and $B_{\rs IB}$ as quantum transitions between the superconductor and the BM and the BM and the superinsulator, respectively, and the range of fields $B_{\rs SB}<B<B_{\rs IB}$ as the domain of the existence of the Bose metal, i.e. the bosonic topological insulator. The intermediate crossing point at $B=B_{\rs SI}$ is a remnant of the SIT tri-critical point. One immediately observes that the expected duality relations $B_{\rs SB}/B_{\rs SI}=B_{\rs SI}/B_{\rs IB}$ and $R_{\rs SB}/R_{\rs SI}=R_{\rs SI}/R_{\rs IB}$, are satisfied with fantastic accuracy. This accuracy is an additional indication of the topological nature of the anomalous metal appearing between the insulating and superconducting phases at the SIT.
Generalizing the standard SIT scaling considerations of\,\cite{fisher1992} onto the BKT transition, we introduce the scaling variable $|g -g_c|({\rm ln} (T_0/T))^2$, where $T_0$ is the adjusting parameter to be determined by the best fit. One expects BKT scaling behaviors near both quantum critical points $B_{\rs SB}$ and $B_{\rs IB}$. Furthermore, since the observed SIT physics is dominated by the proximity to the quantum tri-critical point at $\eta = 1$, one may expect the remnant of the analogous behavior at $B_{\rs SI}$. This calls for revisiting the original scaling result of\,\cite{fisher1992} as well as the observed persistent scaling in the experiment\,\cite{marcus}.
The results of the BKT scaling analysis around $B_{\rs SB}$,  $B_{\rs IB}$, and $B_{\rs SI}$ are presented in Fig.\,\ref{Fig3}d,e,f respectively, and demonstrate an excellent critical fit, supporting the quantum BKT nature of the transitions to the Bose metal phase. Note that, while for both the superconductor- and insulator-to-Bose metal transitions the best fit yields the characteristic temperature $T_0=2$\,K, the corresponding characteristic parameter for the SIT is $T_0=1$\,K, which reflects the different energy characteristics for the remnant quantum SIT occurring near the tri-critical point $g=1$,\,$\eta=1$.

Shown in Fig.\,\ref{Fig4}a are the magnified $R_{\rs \square}(T)$ dependencies which we re-plotted as functions of $1/T$ for the fields below 0.16\,T. We present a few representative curves to avoid data crowding. At relatively high temperatures one sees the resistance rapidly dropping as function of $1/T$ due to thermally activated vortex motion, $R_{\rs \square}(1/T)\propto\exp(-T^*/T)$, the exponential behavior is shown by dashed line. At fields, $B<B_{\rs SB}$, the resistance $R_{\rs \square}(T)$ saturates at low temperatures, indicating the possible crossover to quantum vortex creep. Our findings are in accord with the recently reported dissipative state with non-zero resistance in two-dimensional 2H-NbSe$_2$ films\,\cite{Benyamini2019}.
Above $B_{\rs SB}= 0.011$\,T the $R_{\rs \square}(1/T)$ dependencies develop minima, which become less pronounced above the $B_{\rs SI}$ field, where the insulating behavior becomes dominant. These emergent minima signal that the bulk spectrum of charge excitations  acquires a gap, which prevents bulk electronic transport. Such minima are often viewed as the hallmark of topological insulators, see, for example,\,\cite{Wang2014} and references therein. We now note that in the BM, i.e. topological insulator domain, $B_{\rs SB}<B<B_{\rs IB}$, one can neglect the contribution from moving vortices as it is seen from Fig.\,\ref{Fig4}a. Then the sheet resistance of the BM results from the charge current contribution from two parallel channels, (i)\,the ballistic edge modes and (ii)\,the thermally activated bulk modes over the CS gap. 
Accordingly, below the minimum, $T<T_{\mathrm{min}}$, the sheet resistance is perfectly fitted by the two parallel resistors formula $R_{\rs \square}(T,B)=R_{\rs CS}(T)R_{\mathrm{bal}}(B)/[R_{\rs CS}(T)+R_{\mathrm{bal}}(B)]$, where $R_{\rs CS}(T)\propto\exp(T_{\rs CS}/T)$ is the bulk thermally activated resistance corresponding to surmounting the insulating gap $T_{\rs CS}$, while $R_{\mathrm{bal}}(B)$ is the field dependent resistance mediated by the edge modes. A fit for the field 0.05\,T is shown by the dashed line in\, Fig.\,\ref{Fig4}a.  Upon increasing the magnetic field above $B_{\rs SI}=0.039$\,T (i.e. moving into the $g<1$ region), $R_{\mathrm{bal}}(B)$ increases and the edge states incrementally mix with the bulk modes. As a result, the minimum in $R_{\rs \square}(T,B)$ becomes less pronounced and the film crosses over continuously to insulating behavior. 
Markedly, the $R(T)$ vs.$1/T$ dependence in the TDBG exhibits remarkable similarity to that in NbTiN, see Appendix.

The correlation lengths $\xi_{\rs SB}$$\sim$$1/T_{\mathrm{min}}$ and $\xi_{\rs IB}$$\sim$$1/T_{\mathrm{dev}}$ associated with the quantum BKT superconductor-BM and superinsulator-BM transitions are expected to display the BKT criticality. Here $T_{\mathrm{min}}$ is the position of the minimum in $R_{\rs\square}$, heralding the emergence of the bulk CS insulating gap. The temperature $T_{\mathrm{dev}}$, identified by the deviation of $R_{\rs\square}(T)$ in Fig.\,\ref{Fig4}a from the insulating exponential dependence $\exp(\mathrm{const}/T)$, marks the switching on of the edge modes and the start of shunting  the insulating bulk by metallic edge channels. The latter can be viewed as quantum wires, hence their switching on can be described as a BKT quantum phase transition in analogy to the seminal work\,\cite{Zaikin1992}. Scaling of $1/T_{\mathrm{min}}$ as function of the dimensionless tuning parameter $B/B_{\rs SI}$ is presented in\,Fig.\,\ref{Fig4}b by green solid circles. Accordingly, the positions of the minima of $R(T)$ of TDBG at different $B_{\rs{||}}$ are shown by violet diamonds. Also displayed in Fig.\,\ref{Fig4}b are similar dependencies for other materials, metallic granular films\,\cite{jaeger} and JJA\,\cite{Zant1992}. The data comply perfectly with the BKT exponential scaling illustrating the universality of the transition into the BM in different systems. 
The scaling of $1/T_{\mathrm{dev}}$ is shown in Fig.\,\ref{Fig4}c, although there the window of the magnetic fields is less wide and the error-bar near $B=B_{\rs IB}$ grows large. Note that the exponential BKT scaling can be viewed as a result of the formal $\nu\to\infty$ limit. An ``infinite" critical exponent $\nu$ implies, by the Harris criterion\,\cite{harris}, that disorder is irrelevant for the SIT in the renormalization group sense. The role of disorder in the large-scale properties of the system is only to tune the SIT and renormalize material parameters.

The presented gauge theory of the Bose metal reveals that the long-debated SIT-intervening metallic phase is a bosonic topological insulator. Its metallic conductance is mediated by bosonic edge modes and the superconductor-topological insulator and topological insulator-superinsulator transitions are quantum BKT phase transitions. The BKT scaling at the transition points, together with the high-precision duality relations for the transition fields and the corresponding resistances, providing an unambiguous evidence for the bosonic topological insulator and the associated quantum BKT transitions, are reported for the first time.  Our observation of a bosonic topological insulator in NbTiN and similar behavior in TDBG,  stresses its universal character as a result of the Chern-Simons gap in the spectrum of relevant excitations. Note that the occurrence of bosonic topological insulator in the double belayered graphene  is in a concert with the expectations of\,\cite{Balents2017}. 
Interestingly, our previous findings of\,\cite{Mironov2018} support also the percolation network picture\,\cite{Chalker} of gapless bosonic channels near the BKT transition to TI. Finally, our results resolve the long-standing puzzle of the SIT,  unraveling that the system follows an indirect transition scenario through the intervening metallic state in systems with strong quantum fluctuations, while films with suppressed quantum fluctuations (more disordered and/or with the higher carrier densities) exhibit the direct SIT.

Recent transport studies\,\cite{Kapitulnik2017, Shahar2018} reported that this anomalous Bose metal state possesses neither Hall resistance nor cyclotron resonance, which are expected to be pronounced in states that have longitudinal resistance much smaller than the normal resistance of the respective materials. These findings comply with our prediction that the Bose metal is a bosonic topological insulator. More experimental research, however, is required for conclusive evidence. One can also propose an imaging study of the fluctuation pattern using local probes, like scanning squid, scanning NV microscopy, or scanning impedance microscopy, as well as using a non-local measurement geometry to obtain indirect evidences of the edge modes.


\subsection*{Acknowledgments}
We
are delighted to thank Thomas Proslier for preparing NbTiN samples used in the experiments 
and 
Tom Rosenbaum for valuable discussions. M.C.D. thanks CERN, where she completed this work, for kind hospitality. S.V.P. is grateful to Tatyana Baturina for valuable contribution at the initial stage of work  on  NbTiN films.
The work at Argonne (V.M.V.) was supported by the U.S. Department of Energy, Office of Science, Basic Energy Sciences, Materials Sciences and Engineering Division.
The work on transport measurements at Novosibirsk was supported by the grant of RF president (MK-5455.2018.2). The work by Y.K. was supported by FAPESP, CNPq and AFOSR Grant FA9550-17-1-0132. The work at Caltech (D.S.) was supported by National Science Foundation Grant No. DMR-1606858.

\section*{Appendix}
		\subsection{Free energy}~~
	We start with the
	the action describing the intertwined vortex-CP dynamics derived in\,\cite{dtv}:
	\begin{eqnarray}
	\nonumber S = \int dt d^2 x\,\left[  i \frac{\bar{q}}{ 2\pi} a_{\mu} \epsilon_{\mu \alpha \nu} \partial_{\alpha} b_{\nu}
	+ \frac{v_c}{ 2e^2_{\mathrm v} } f_0 f_0+ \frac{1}{ 2e^2_{\mathrm v}v_c} f_i f_i \right.  \\
	+\frac{v_c}{2e^2_{\mathrm q}} g_0 g_0 +\frac{1}{2e^2_{\mathrm q}v_c} g_i g_i
	\left.+ i\sqrt{\bar{q}} a_{\mu} Q_{\mu} +i\sqrt{\bar{q}} b_{\mu} M_{\mu}\right]\,,
	\label{nonrel}
	\end{eqnarray}
	where  $v_{\mathrm c} = 1/\sqrt{\mu \varepsilon}$ is the speed of light in the film material, expressed in terms of the magnetic permeability $\mu$ and  the electric permittivity $\varepsilon$ (we use natural units $c=1$, $\hbar = 1$).
	This action (\ref{nonrel}) is a non-relativistic version of the topologically massive gauge theory\,\cite{jackiw} describing a (2+1)-dimensional vector particle with the CS mass, $m_{\rs CS}=\bar{q} e_{\mathrm q} e_{\mathrm v}/2\pi v_c$, arising without spontaneous symmetry. The emergent gauge fields $a_{\mu}$ and $b_{\mu}$ mediate the mutual statistics interactions between Cooper pairs, with word-lines $Q_{\mu}$ and charge $\bar{q} = 2$ and vortices of flux $2\pi/\bar{q}$, with world-lines $M_{\mu}$.
	When appropriately regularized on a lattice of spacing $\ell$ (see Supplementary Information), these world-lines can be viewed as ``strings" of typical length $L=N\ell$, carrying electric and magnetic quantum numbers $Q$ and $M$, respectively.
	To derive the free energy of the interacting Cooper pair-vortex system the gauge fields in the above quadratic action are integrated out
	via the standard Gaussian integration procedure, obtaining thus an effective action for the charge and vortex strings alone. As usual in statistical field theory this has the interpretation of an energy for the ``string gas", which is proportional to the string length.
	To obtain a free energy associated with the strings, one has to include the contribution from the positional string entropy, which is also proportional to its length, with the proportionality factor $\mu_{\mathrm e} = {\rm ln} (5)$ representing the 5 possible choices for string continuation at each lattice site. For the relevant case of Cooper pairs (i.e. $\bar{q} =2$), we find the free energy in the main text.

	\subsection{Effective action for the topological insulator}~~
	To determine the nature of the Bose metal we find its electromagnetic response by coupling the charge current $(\bar q e) j_{\mu}$
	to an external electromagentic potential $A_{\mu}$ and we compute its effective action by integrating out gauge fields $a_{\mu}$ and $b_{\mu}$,
	\begin{eqnarray}
	{\rm e}^{-S_{\rm eff} \left( A_{\mu} \right) }= \frac{1}{Z} \ \int {\cal D} a_{\mu} {\cal D} b_{\mu} {\rm e}^{-S \left( a_{\mu}, b_{\mu} \right) + i (\bar{q} e) j_{\mu} A_{\mu }  } \ ,
	\nonumber \\
	Z = \int {\cal D} a_{\mu} {\cal D} b_{\mu} {\rm e}^{-S \left( a_{\mu}, b_{\mu} \right)  }\,.
	\label{integration}
	\end{eqnarray}
	This gives
	\begin{equation}
	S_{\rm eff} \left( A_{\mu} \right) = \frac{g }{ 4} \left( \frac{\bar{q} e}{2\pi} \right)^2 d \int d^3 x \left( v_c F_0^2 + {1\over v_c} F_i^2 \right) \ ,
	\label{effac}
	\end{equation}
	where $F_{\mu}$$=$$\epsilon_{\mu \alpha \nu} \partial_{\alpha} A_{\nu}$ is the dual field strength and we have identified the geometric lattice factor $4 \mu_ e \eta \ell$ with the relevant thickness parameter $d$ of the film, so as to maintain self-duality (see main text).
	This is the action of a bulk insulator which becomes best evident in the relativistic case, $v_c=1$:
	\begin{equation}
	S_{\rm eff} \left( A_{\mu} \right) = \frac{g }{2} \left( \frac{\bar{q} e}{2\pi} \right)^2 d \int d^3 x \ A_{\mu} \left( -\delta_{\mu \nu} \nabla^2 + \partial_{\mu} \partial_{\nu} \right) A_{\nu} \,.
	\label{bulkaction}
	\end{equation}
	Varying this action with respect to the vector potential $A_\nu$ gives the main text formula for the electric current.

	\subsection{Conduction by edge modes}
	The Chern-Simons effective action is not invariant under gauge transformations $a_i=\partial_i \lambda$ and $b_i=\partial_i \chi$ at the edges. Two chiral bosons\,\cite{flore} $\lambda = \xi + \eta$ and $\chi = \xi -\eta$ have to be introduced to restore the full gauge invariance, exactly as it is done in the quantum Hall effect framework\,\cite{wen} and for topological insulators\,\cite{moore}. The full gauge invariance is restored by adding the edge action
	\begin{eqnarray}
	S_{\rm edge} = {1 \over \pi} \int d^2 x \ \left( \partial_0 \xi \partial_s \xi  - \partial_0 \eta \partial_s \eta \right) + \nonumber \\
	\bar{q} e_{\rm eff}\int d^2 x \  A_0  \left( {\sqrt{\bar{q}}\over 2\pi}\partial_s \chi \right) \ ,
	\label{edgeac}
	\end{eqnarray}
	including the electromagnetic coupling of the edge charge density $\rho= (\bar{q} e_{\rm eff}) (\sqrt{\bar{q}}/ 2\pi)\partial_s \chi $
	in the $A_s=0$ gauge, $\bar{q} e_{\rm eff}$ being the effective charge of the Cooper pairs in the Bose metal phase, $\bar{q} e_{\rm eff} = \bar{q} e \sqrt{g}$.
	As in the case of the quantum Hall effect, the non-universal dynamics of the edge modes is generated by boundary effects \cite{wen}, which result in the Hamiltonian
	\begin{equation}
	H = {1 \over \pi} \int ds \left[ -v_b \left( \partial_s \xi \right)^2 -v_b \left( \partial_s \eta \right)^2\right] \ ,
	\label{four}
	\end{equation}
	where $v_b$ is the velocity of propagation of the edge modes along the boundary. Upon adding this term, the total edge action becomes
	\begin{eqnarray}
	S_{\rm edge} = {1 \over \pi} \int d^2 x \ \left[ \left(\partial_0-v_b\partial_s\right) \xi \partial_s \xi  - \left(\partial_0+v_b \partial_s \right)\eta \partial_s \eta \right]\nonumber \\ + \bar{q} e_{\rm eff} \int d^2 x \  A_0  \left( {\sqrt{\bar{q}}\over 2\pi}\partial_s \chi \right) \ .
	\label{edgeacfull}
	\end{eqnarray}
	The equation of motion generated by this action is
	\begin{equation}
	v_b \partial_s \rho= {\bar{q} e_{\rm eff} \over 2\pi} E = {\bar{q} e_{\rm eff} \over 2\pi} \partial_s A_0\ .
	\label{eight}
	\end{equation}
	Integrating this equation gives eq. (\ref{response}) in the main text, which represent ballistic charge conduction with the resistance $R =R_{\rm Q}/g$.

	\subsection{Bose metal stability}~~
	To analyze an intervening phase harboring dilute topological excitations in the~Hamiltonian~formalism, we set $Q_{\mu }$$=$$M_{\mu }$$=$$0$ and decompose the original gauge fields in Eq.\,(\ref{nonrel}) as $a^i $$=$$ \partial_i \xi $$+$$\epsilon^{ij} \partial_j \phi $,\,$b^i$$=$$\partial_i \lambda$$+$$\epsilon^{ij} \partial_j \psi $. Quantizing the action in~(\ref{nonrel})
	we arrive at the ground state wave functional
	\begin{widetext}
	$\Psi [a^i,b^i] $$=$$\exp \left[ { i (\bar{q}/4 \pi)} \int d^2{\bf x} \left( \psi \Delta \xi + \phi \Delta \lambda \right) - (\bar{q}/4\pi)\int d^2{\bf x} \left( g(\partial_i \phi)^2 +  {1\over g} (\partial_i \psi)^2\right)  \right]$,
	\end{widetext}
	generalizing the  Schr\"{o}dinger wave function to a system with an infinite number of degrees of freedom. The fields $\phi $ and $\psi$ represent vortex- and CP charge-density waves, which are gapped due to the mutual statistics interactions.
	When the two symmetries are compact, however, the fields $\xi$ and $\lambda$ are angles and we have to take into account also the corresponding topological excitations, vortices and point charges. These can be in highly entangled mixed states or in their pure free state. Quantum operator expectation values in the entangled mixed state in which vortices are the non-observed environment, are given by 
	\begin{widetext}$\langle{\cal O}\rangle$$\propto$$\int {\cal D}\psi {\cal D}\lambda\,{\cal O}(\psi, \lambda)\
	{\rm exp} \left(-\int d^2{\bf x} \ {{\bar q}  \over 2\pi g} \left( \partial_i \psi \right)^2 + {{\bar q} \over 2\pi g} \left( \partial_i \lambda\right)^2
	- 2z {\rm cos} \lambda \right)$, 
	\end{widetext}
see SI. The quantum fugacity $z$ governs the degree of entanglement. This is the classical partition function of the sine-Gordon model undergoing the BKT transition\,\cite{kos, ber}. In a quantum case, it is a quantum BKT transition at the ``effective temperature" for vortex liberation set by the quantum conductance parameter $g$. Correspondingly, charge liberation from the dual entangled state is set by $1/g$. The highly entangled states correspond to superconductor and superinsulator at high and low values of $g$, respectively, where  vortices and charges have algebraic correlation functions. The Bose topological insulator is the intervening state at  $g\simeq 1$, where both charges and vortices are screened by strong quantum fluctuations, leading to correlation functions (\ref{corrbkt}).

	\subsection{Samples and measurements}
	
	To grow NbTiN films, we employed the atomic layer deposition (ALD) technique based on sequential surface reaction step-by-step film growth. The fabrication technique is described in detail in the Supplemental Material. This highly controllable process provides superior thickness and stoichiometric uniformity and an atomically smooth surface\,\cite{Lim:2003}
	as compared to chemical vapor deposition, the standard technique used to grow NbTiN films.
	We used NbCl$_5$, TiCl$_4$, and NH$_3$ as gaseous reactants; the stoichiometry was tuned by varying the ratio of TiCl$_4$/NbCl$_5$ cycles during growth\,\cite{Proslier:2011}. The superconducting properties of these ultrathin NbTiN films were optimized by utilizing AlN buffer layers grown on top of the Si substrate\,\cite{Shiino:2010}. Nb$_{1-x}$Ti$_x$N films of thicknesses $d=10$ were grown. Films have a fine-dispersed polycrystalline structure~\cite{Mironov2018}. The average crystallite size is $\approx 5$\,nm. Deposition temperature is 350$^0$C. Ti fraction $x$ is 0.3.
	
	The films were lithographically patterned into bridges 50 $\mu$m wide, the distance between current-contacts was 2500 $\mu$m and distance between voltage-contacts was 450 $\mu$m. Most resistive transport measurements are carried out using low-frequency ac techniques in a two-terminal configuration with $V \approx 100$ $\mu$V, $f \approx 1$ Hz. Additionally we measured temperature dependence of resistance at zero magnetic field. From comparison two- and four-terminal configuration we determined number of square in a two-terminal configuration for obtaining resistance per square. For ac measurements we use SR830 Lock-ins and current preamplifiers SR570. All the resistance measurement are carried out in linear regime with using adequately system of filtration. Resistivity measurements at sub-Kelvin temperatures were performed in dilution refrigerators $^3$He/$^4$He with superconducting magnet.
	
	\begin{widetext}
	\subsection{Low-temperature resistance in NbTiN and double belayer graphene}~~\\
	\vspace{-1cm}
	\begin{figure*}[b!]
		\vspace{-1cm}
		\centering
		\includegraphics[width=5cm]{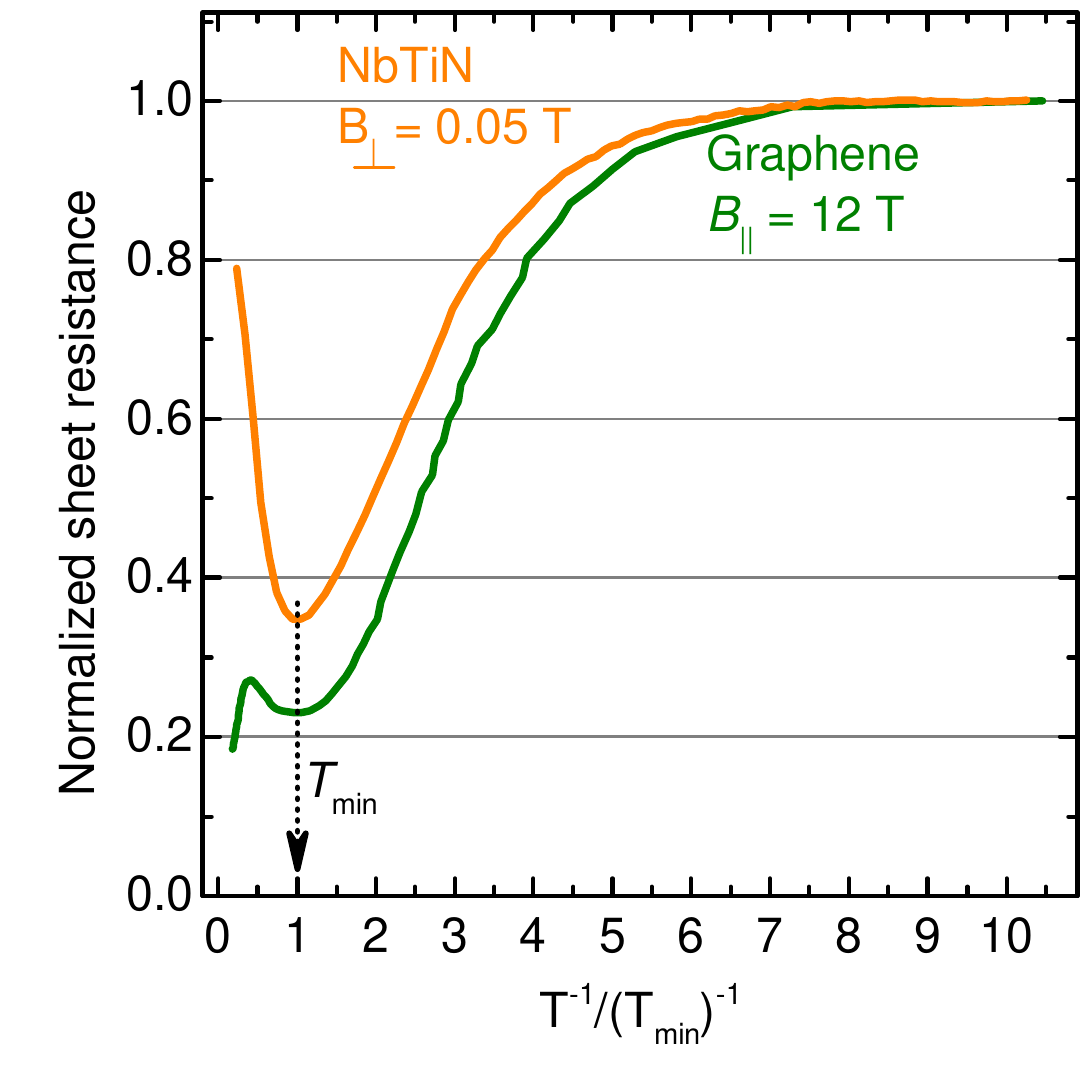}
		\vspace{-0.5cm}
		\caption{\textbf{Sheet resistance vs $1/T$ plots in NbTiN and twisted double belayer graphene (TDBG)}
			Representative $R_{\rs \square}$ vs. $1/T$ plots for fields $B_{\rs\perp}=0.05$\,T for NbTiN and $B_{\rs{||}}=0.05$\, for TDBG (sample No.\,2, see main text). The temperature scales are normalized with respect to positions of the minima, resistances are normalized with respect to their saturation values. Importantly, normalized temperatures of the saturation coincide for both systems indicating a universal topological character of the quantum BKT transitions confining the domain of the existence of the bosonic topological insulators.
		}
		\label{Fig5}
	\end{figure*}
\end{widetext}

%


\end{document}